\documentclass[conference,a4paper]{IEEEtran}
\usepackage{cite}
\usepackage{amsmath,amssymb,amsfonts}
\usepackage{algorithmicx}
\usepackage{algorithm}
\usepackage[noend]{algpseudocode}
\makeatletter
\def\BState{\State\hskip-\ALG@thistlm}
\usepackage{graphicx}
\usepackage{textcomp}
\usepackage{xcolor}
\usepackage{setspace}
\def\BibTeX{{\rm B\kern-.05em{\sc i\kern-.025em b}\kern-.08em
		T\kern-.1667em\lower.7ex\hbox{E}\kern-.125emX}}
\usepackage{subfigure}
\newtheorem{my_theorem}{Theorem}

\newtheorem{my_corollary}{Corollary}
\newtheorem{my_proposition}{Proposition}

\makeatletter
\def\blfootnote{\xdef\@thefnmark{}\@footnotetext}
\makeatother
\begin{document}
	\title{A Generalized Statistical Model for THz wireless Channel with Random Atmospheric Absorption}
	
	\author{
		\IEEEauthorblockN{Pranay Bhardwaj and S. M. Zafaruddin}\\
		\IEEEauthorblockA{ Deptt. of Electrical and Electronics Engineering, Birla Institute of Technology and Science, Pilani, Rajasthan, India.\\ Email: \{p20200026, syed.zafaruddin\}@pilani.bits-pilani.ac.in}
		
		\thanks{  }
	}
\thispagestyle{empty} 	
	\maketitle
\begin{abstract}
Current statistical channel models for Terahertz (THz) wireless communication primarily concentrate on the sub-THz band, mostly with $\alpha$-$\mu$ and Gaussian mixture fading distributions for short-term fading and deterministic modeling for atmospheric absorption. In this paper, we develop a  generalized statistical model for signal propagation at THz frequencies considering random path-loss employing  Gamma distribution for the molecular absorption coefficient, short-term fading characterized by the $\alpha$-$\eta$-$\kappa$-$\mu$ distribution, antenna misalignment errors, and transceiver hardware impairments. The proposed model can handle various propagation scenarios, including indoor and outdoor environments, backhaul/fronthaul situations, and complex urban settings. Using Fox's H-functions, we present the probability density function (PDF) and cumulative distribution function (CDF) that capture the combined statistical effects of channel impairments. We analyze the outage probability of a THz link to demonstrate the analytical tractability of the proposed generalized model. We present computer simulations to demonstrate the efficacy of the proposed model for performance assessment with the statistical effect of atmospheric absorption. 
\end{abstract}

\begin{IEEEkeywords}
Antenna misalignment error, atmospheric absorption,	channel model, hardware impairment, THz wireless systems. 
\end{IEEEkeywords}

\blfootnote{This work was supported in part by the Science and Engineering Research Board (SERB), Government of India, under MATRICS Grant MTR/2021/000890.}

\section{Introduction} \label{sec:intro}
Terahertz (THz) communication is gaining rapid momentum as a promising technology for the upcoming generation of wireless networks, often referred to as 6G. With its generous unlicensed bandwidth spanning from $0.1$ to $10$ THz, the THz band enables the delivery of high data rates while simultaneously ensuring low latency and enhanced security for backhaul/fronthaul communications \cite{Koenig_2013_nature,Dang_2020_nature}. Additionally, THz wireless technology has the potential to play a crucial role for both backhaul and access networks, particularly in dense network deployments with limited coverage range \cite{Song_2011_THz,Sarieddeen_2020_THz_survey}. A generalized channel model specifically for the THz band is desirable, considering its distinguishing channel characteristics compared with the mmWave and radio-frequency (RF) technologies.

Despite being a promising technology, THz wireless transmission encounters several challenges, including high path-loss resulting from molecular absorption, misalignment errors, short-term fading, and hardware impairments. The THz band experiences higher path-loss due to signal absorption by molecules at extremely small wavelengths \cite{Kim2015, Kokkoniemi_2018, Wu2020}. In \cite{Kim2015}, the authors experimentally characterized the THz channel (300-320 GHz). The impact of scattering and absorption losses on THz wavelengths, particularly in the absence of white noise, was investigated in \cite{Kokkoniemi_2018}. The study presented in \cite{Wu2020} analyzed signal propagation in indoor environments, considering blockage effects caused by walls and human bodies. The effect of antenna misalignment errors is also detrimental to THz performance, which occurs when transmit and receive antennas fail to adequately align for line-of-sight (LOS) transmissions, significantly limiting the physical communication range. Recently, a statistical model for antenna misalignment errors for the THz system appeared in \cite{Dabiri2022_zaf}. As is for many wireless communications, short-term fading is inevitable at THz frequencies. Several studies have suggested various distributions to model the short-term fading in the THz band, like $\alpha$-$\mu$, fluctuating two-ray (FTR), and Gaussian mixture distributions  at sub-THz frequencies \cite{Papasotiriou2021_scientific_report}\cite{Papasotiriou2023_scintific_report_outdoor}. Notably, in indoor environments at $ 142 $ \mbox{GHz}, experiments in \cite{Papasotiriou2021_scientific_report} confirmed the suitability of the $\alpha$-$\mu$ distribution, while in outdoor settings, \cite{Papasotiriou2023_scintific_report_outdoor} recommended the Gaussian mixture distribution for efficient characterization of short-term fading. The achievable performance can be limited by transceiver hardware impairment at higher frequencies. The statistical modeling of residual hardware impairment has employed the Gaussian distribution \cite{Schenk_hw}. Extensive performance analysis has been conducted for a sub-THz system, considering both antenna misalignment errors and short-term fading. The studies were carried out in two scenarios: one without hardware impairments, as reported in \cite{Pranay_2021_TVT, Pai2021_dual_hop_THz_backhaul, Li_2021_THz_AF, Bhardwaj2022_multihop, Bhardwaj2022_systems_journal, Joshi2022_ftr, Bhardwaj2023_outdoor_thz, Bhardwaj2023_iot}, and the other with the consideration of hardware impairments \cite{Bhardwaj2023_THI, Boulogeorgos_Analytical}.

All the previously mentioned research has focused on sub-THz frequencies (primarily within the $150$ \mbox{GHz} range). In this context, the statistical model relies on a short-term model with moderate flexibility, parameters, and deterministic path loss caused by atmospheric absorption. At higher, true THz frequencies, the suitability of the proposed $\alpha$-$\mu$ and Gaussian mixture fading distributions might not accurately represent the majority of propagation scenarios. Furthermore, the interaction of THz signals with the atmosphere can become intricate, necessitating complex modeling for atmospheric absorption.   Thus, developing a generalized channel model employing random atmospheric absorption and a multi-parameter fading model with the statistical effect of antenna misalignment and hardware impairment is desirable to accommodate a broader range of propagation scenarios at THz frequencies for a better performance estimate.

In this paper, we develop a generalized statistical model for THz signal propagation, which considers various factors such as random path-loss using the Gamma distribution for atmospheric absorption, short-term fading characterized by the $\alpha$-$\eta$-$\kappa$-$\mu$ distribution, antenna misalignment errors, and transceiver hardware impairments. The proposed model is highly versatile, accommodating diverse propagation scenarios, including indoor and outdoor environments, backhaul/fronthaul setups, and complex urban settings with obstacles. This analysis helps understand the system's performance under different conditions and highlights the effectiveness of the generalized model in diverse THz communication scenarios. We present the probability density function (PDF) and cumulative distribution function (CDF) to capture the combined statistical effects of channel impairments in terms of Fox's H-functions. Moreover, we analyze the outage probability of a THz link, demonstrating the analytical tractability of the proposed generalized model. To validate the effectiveness of the model in performance assessment, computer simulations are carried out, considering the statistical effect of atmospheric absorption. 

\linespread{0.88}
	
\section{System Model}
We consider a system where a single user transmits a  signal from source to destination within the THz band. The received signal $y$ at the destination is given by \cite{Schenk_hw}
\begin{flalign} \label{eq:rec_sig}
	y = h(s+w_t) + w_r +w
\end{flalign}
where $h$ represents the channel coefficient, comprising of the path gain $h_l$ due to signal propagation and atmospheric absorption, short-term fading $h_f$, and antenna misalignment errors $h_p$. Here, $s$ denotes the desired signal, and $w$ stands for the additive white Gaussian noise (AWGN) with variance $\sigma_\omega^2$. The terms $w_t$ and $w_r$ refer to components representing residual hardware impairment, which are statistically modeled using a Gaussian distribution \cite{Schenk_hw,Boulogeorgos_Analytical}. Specifically, $w_{t}\sim\mathcal{CN}(0,k_{t}^{2}P)$, and $w_{r}\sim\mathcal{CN}(0,k_{r}^{2}P\lvert h \rvert^{2})$, where $k_{t}$ and $k_{r}$ quantify the extent of hardware imperfections in the transmitter and receiver, respectively. In the THz band, typical values of $k_t$ and $k_r$ fall within the range of $(0$-$0.4)$, as suggested in \cite{Boulogeorgos_Analytical}. A value of $k_t=k_r=0$ corresponds to an ideal front-end, representing no hardware imperfections.

We define the instantaneous SNR of the THz link at the receiver as $\gamma = |h|^2 \bar{\gamma}$, where $h = h_lh_fh_p$ and $\bar{\gamma}$ is the average SNR of the link and is defined as $\bar{\gamma} = \frac{P_t}{\sigma_\omega^2}$ with $P_t$ as the transmit power. From \eqref{eq:rec_sig}, the resultant SNR with transceiver hardware impairment is given by
\begin{flalign}
\gamma = \frac{\bar{\gamma} |h|^2}{k_h^2 \bar{\gamma} |h|^2 +1}
\end{flalign}

To model the short-term fading in the THz band for outdoor environment, we adopt the $\alpha$-$\eta$-$\kappa$-$\mu$   fading model with PDF of the channel envelope given by \cite{Bhardwaj2023_alpha_eta_kappa_mu_globecom}
\begin{flalign} \label{eq:pdf_aekm}
	&f_{h_f}(h_f) = \frac{\psi_1 \pi^2 2^{(2-\mu)} A_4^{A_1} A_5^{A_2} h_f^{\alpha\mu-1} e^{-\psi_3 h_f^\alpha}}{(\hat{r}^\alpha)^{1+\frac{\mu}{2}}} \nonumber \\ &\times H^{0,1;1,0;1,1;1,0}_{1,1;0,1;2,3;1,3} \Bigg[ \begin{matrix}~ V_1~ \\ ~V_2~ \end{matrix} \Bigg| A_{3}h_f^{\alpha}, \frac{A_4^2}{4}h_f^{\alpha},\frac{A_5^2}{4}h_f^{\alpha}\Bigg],
\end{flalign}
where  $V_1 = \big\{(-A_2;1,0,1)\big\}: \big\{(-,-)\big\} ; \big\{(-A_1,1)(\frac{1}{2},1)\big\} ;\big\{(\frac{1}{2},1)\big\} $ and $V_2 = \big\{(-1-A_1 -A_2;1,1,1) \big\} : \big\{(0,1) \big\} ; \big\{(0,1),(-A_1,1),(\frac{1}{2},1) \big\} ; \big\{(0,1),(-A_2,1),(\frac{1}{2},1) \big\}$, and the constants are defined as $ \psi_1 = \frac{p\alpha\mu^{2}\xi^{1+\frac{\mu}{2}}\delta^{\frac{\mu}{2}-1}q^{\frac{1+p-p\mu}{2+2p}}\eta^{-\frac{1+p+p\mu}{2+2p}}}{\kappa^{\frac{\mu}{2}-1}\exp \left (\frac{(1+pq)\kappa\mu}{\delta}\right)}$, $ \psi_2$ = $\alpha-1$, $ \psi_3= \frac{p\xi\mu}{\eta\hat{r}^\alpha} $, $ A_{1} = \frac{p\mu}{1+p}$-$1, A_{2}= \frac{\mu}{1+p}$-$1, A_{3}= \frac{(\eta-p)\xi\mu}{\eta\hat{r}^\alpha} $, $ A_{4} = 2p\mu \sqrt{\frac{q\kappa\xi}{\eta\delta\hat{r}^\alpha}} $, and $A_{5} = 2\mu\sqrt{\frac{\kappa\xi}{\delta\hat{r}^\alpha}} $. Here, $\alpha$ represents the nonlinearity characteristic of the medium, while $\eta$ denotes the ratio of the total power of in-phase and quadrature scattered waves within the multipath clusters. Furthermore, $\kappa$ is defined as the ratio between the total power of dominant components and the total power of scattered waves, and $\mu$ stands for the number of multipath clusters present. 

The antenna misalignment errors in the THz band \cite{Dabiri2022_zaf, Badarneh2023_THz_Pointing} are statistically modeled using the PDF
\begin{flalign} \label{eq:pdf_pointing_thz}
f_{h_{p}}^{}(h_p) = -\rho^2 \ln(h_p) h_p^{\rho-1}
\end{flalign}
where $0<x<1$. $\rho$ determines the severity of the misalignment errors. 

\section{Random Path Loss}
This section focuses on exploring the path loss modeling in the THz link and utilizing it to develop a generalized statistical model. The path gain, denoted as $h_l$, relies on various factors such as antenna gains, frequency, and molecular absorption coefficient, as defined in \cite{Boulogeorgos_Analytical}:
\begin{flalign} \label{eq:det_path_loss_eqn}
h_l = \frac{c\sqrt{G_{t}G_{r}}}{4\pi f d} \exp\bigg(-\frac{1}{2}\zeta(f,T,\psi,p)d\bigg)
\end{flalign}
Here, $c$ represents the speed of light, $f$ denotes the transmission frequency, and $d$ represents the distance in \mbox{m}. $G_{t}$ and $G_{r}$ are the antenna gains of the transmitting and receiving antennas, respectively. The term $\zeta(f,T,\psi,p)$ corresponds to the molecular absorption coefficient, which is influenced by temperature $T$, relative humidity $\psi$, and atmospheric pressure $p$.  At sub-THz frequencies, $\zeta(f,T,\psi,p)$ has been considered deterministic. However, at higher frequencies, the interaction of THz signals with the atmosphere can become intricate at the molecular level, making it essential to employ complex modeling for the molecular absorption coefficient $\zeta(f, T, \psi, p)$.

As measurement data beyond a few hundred \mbox{GHz} is unavailable for parameterizing $\zeta(f, T, \psi, p)$ at THz frequencies, we resort to a statistical approach. We adopt the Gamma distribution $f_{\zeta(f, T, \psi, p)}(x) = \frac{x^{k-1}}{\beta^k\Gamma(k)} e^{-\frac{x}{\beta}}$ with parameters $k$ and $\beta$ in dB to model $\zeta(f, T, \psi, p)$. This statistical model allows for a diverse range of support for $\zeta(f, T, \psi, p)$ between $0$ and $\infty$, with an average value of $k\beta$ dB/km selected from existing measurement data. In a recent study, the Gamma distribution has been utilized to model the attenuation coefficient in free-space optics (FSO) transmission under foggy weather conditions \cite{fog}. Indeed, experimental data is crucial to validate the use of the Gamma distribution for the absorption coefficient observed in practical scenarios, which presents an excellent opportunity for further research. 
\begin{my_proposition}
If the molecular absorption coefficient  $\zeta(f, T, \psi, p)$ is Gamma distributed with  parameters $k$ and $\beta$ in dB, then the PDF of the path-gain $h_l$ is given by
\begin{eqnarray} \label{eq:pdf_path_loss}
f_{h_{l}}(h_{l}) = \frac{z^k a_l^{-z}}{\Gamma(k)}\bigg[ln\bigg(\frac{a_l}{h_{l}}\bigg)\bigg]^{k-1} h_{l}^{z-1}
\end{eqnarray}
where  $a_l = \frac{c\sqrt{G_tG_r}}{4\pi f d}$, $0<h_{l}\le a_l$, and $z = 8.686/(\beta d)$.
\end{my_proposition}

\begin{IEEEproof}
Converting $\zeta$ in dB, \eqref{eq:det_path_loss_eqn} can be represented by
\begin{equation}
h_l = \frac{c\sqrt{G_{t}G_{r}}}{4\pi f d} \exp(-\frac{1}{2}\zeta^{\rm dB}(f,T,\psi,p)d^{\rm km}/4.343)
\end{equation}
where $ d^{\rm km} $ is the distance in \mbox{km}. Using $f_{\zeta(f, T, \psi, p)}(x) = \frac{x^{k-1}}{\beta^k\Gamma(k)} e^{-\frac{x}{\beta}}$ and applying standard transformation of random variables, the PDF of $h_l$ is given \eqref{eq:pdf_path_loss}.
\end{IEEEproof}

\section{ Statistical  Channel Model and  Performance Analysis}
In this section, we develop a generalized channel model for the THz transmission, which includes the combined effect of random path loss, antenna misalignment errors, the generalized $\alpha$-$\eta$-$\kappa$-$\mu$ short-term fading, and transceiver hardware impairments. The $\alpha$-$\eta$-$\kappa$-$\mu$ model is a comprehensive representation that embodies a wide range of fading characteristics, including the number of multi-path clusters, power of dominant components, nonlinearity of the propagation medium, and scattering level. This flexibility allows the model to effectively capture and fit measurement data in various propagation scenarios, making it an ideal choice for a more generalized and diverse channel model at high frequencies.

\subsection{ Statistical  Model}
In the following theorem, we present the PDF and CDF of a single link THz transmission combining the effects of short-term fading, antenna misalignment errors, and random path loss with transceiver hardware impairments.
\begin{my_theorem}
If $\gamma_h = \sqrt{\frac{\gamma}{\bar{\gamma}(1-\gamma k_h^2)}}$, then the PDF and CDF of SNR combining the effects of random path loss, antenna misalignment errors, and short-term fading with transceiver hardware impairments for a THz link is given by
\begin{flalign} \label{eq:pdf_aekm_rpl}
	&f_{\gamma}(\gamma) = \frac{ (-1)^{-k}\psi_1 \rho^2  z^k  }{(\hat{r}^\alpha)^{1+\frac{\mu}{2}} } \bigg(\frac{1}{a_l}\bigg)^{\alpha(1+A_1+A_2)+1}  \nonumber \\ &  \frac{1}{2\big(1-\gamma k_h^2\big)\sqrt{\frac{\bar{\gamma} \gamma}{1-\gamma k_h^2}}} \bigg(\frac{\gamma_h}{\bar{\gamma}}\bigg)^{\frac{\alpha(1+A_1+A_2)}{2}} \nonumber \\ &  H^{0,2+k;1,0;1,1;1,0;1,0}_{2+k,2+k;0,1;2,3;1,3;0,1} \Bigg[ \begin{matrix}~ V_5~ \\ ~V_6~ \end{matrix} \Bigg| \frac{A_{3}\gamma_h^{\frac{\alpha}{2}}}{a_l^\alpha \bar{\gamma}^{\frac{\alpha}{2}}}, \frac{A_4^2 \gamma_h^{\frac{\alpha}{2}}}{4 a_l^\alpha \bar{\gamma}^{\frac{\alpha}{2}}},\frac{A_5^2 \gamma_h^{\frac{\alpha}{2}}}{4 a_l^\alpha \bar{\gamma}^{\frac{\alpha}{2}}},\frac{\psi_3\gamma_h^{\frac{\alpha}{2}}} {a_l^\alpha \bar{\gamma}^{\frac{\alpha}{2}}} \Bigg],
\end{flalign}
where $V_5 = \big\{(-A_2;1,0,1,0)\big\}, \big\{\rho-\alpha-\alpha A_1 -\alpha A_2; \alpha, \alpha, \alpha, \alpha\big\}, \big\{z-\alpha-\alpha A_1 -\alpha A_2; \alpha, \alpha, \alpha, \alpha\big\}_k : \big\{(-,-)\big\} ; \big\{(-A_1,1)(\frac{1}{2},1)\big\} ;\big\{(\frac{1}{2},1)\big\}; \big\{-, -\big\} $ and $V_6 = \big\{(-1-A_1 -A_2;1,1,1,0) \big\}, \big\{\rho-1-\alpha-\alpha A_1 -\alpha A_2; \alpha, \alpha, \alpha, \alpha\big\}, \big\{z-1-\alpha-\alpha A_1 -\alpha A_2; \alpha, \alpha, \alpha, \alpha\big\}_k : \big\{(0,1) \big\} ; \big\{(0,1),(-A_1,1),(\frac{1}{2},1) \big\} ; \big\{(0,1),(-A_2,1),(\frac{1}{2},1) \big\} ; \\ \big\{ (0,1)\big\}$.

\begin{flalign} \label{eq:cdf_aekm_rpl}
	&F_{h_{fpl}}(x) = \frac{ \psi_1 \rho^2  z^k  }{(\hat{r}^\alpha)^{1+\frac{\mu}{2}}} \Big(\frac{1}{a_l}\Big)^{\alpha(1+A_1+A_2)+1}  \bigg(\frac{\gamma}{\bar{\gamma}}\bigg)^{\frac{\alpha(1+A_1+A_2) +1}{2}}   \nonumber \\ &  H^{0,3+k;1,0;1,1;1,0;1,0}_{3+k,3+k;0,1;2,3;1,3;0,1} \Bigg[ \begin{matrix}~ V_7~ \\ ~V_8~ \end{matrix} \Bigg| \frac{A_{3}\gamma_h^{\frac{\alpha}{2}}}{a_l^\alpha \bar{\gamma}^{\frac{\alpha}{2}}}, \frac{A_4^2 \gamma_h^{\frac{\alpha}{2}}}{4 a_l^\alpha \bar{\gamma}^{\frac{\alpha}{2}}},\frac{A_5^2 \gamma_h^{\frac{\alpha}{2}}}{4 a_l^\alpha \bar{\gamma}^{\frac{\alpha}{2}}},\frac{\psi_3\gamma_h^{\frac{\alpha}{2}}} {a_l^\alpha \bar{\gamma}^{\frac{\alpha}{2}}} \Bigg],
\end{flalign}

where $V_7 = \big\{(-A_2;1,0,1,0)\big\}, \big\{\rho-\alpha-\alpha A_1 -\alpha A_2; \alpha, \alpha, \alpha, \alpha\big\}, \big\{z-\alpha-\alpha A_1 -\alpha A_2; \alpha, \alpha, \alpha, \alpha\big\}_k, \big\{-\alpha-\alpha A_1 -\alpha A_2; \alpha, \alpha, \alpha, \alpha\big\} : \big\{(-,-)\big\} ; \big\{(-A_1,1)(\frac{1}{2},1)\big\} ;\big\{(\frac{1}{2},1)\big\}; \big\{-, -\big\} $ and $V_8 = \big\{(-1-A_1 -A_2;1,1,1,0) \big\}, \big\{\rho-1-\alpha-\alpha A_1 -\alpha A_2; \alpha, \alpha, \alpha, \alpha\big\}, \big\{z-1-\alpha-\alpha A_1 -\alpha A_2; \alpha, \alpha, \alpha, \alpha\big\}_k, \big\{-1-\alpha-\alpha A_1 -\alpha A_2; \alpha, \alpha, \alpha, \alpha\big\} : \big\{(0,1) \big\} ; \big\{(0,1),(-A_1,1),(\frac{1}{2},1) \big\} ; \big\{(0,1),(-A_2,1),(\frac{1}{2},1) \big\} ; \\ \big\{ (0,1)\big\}$.
\end{my_theorem}

\begin{IEEEproof}
	The proof is presented in Appendix A.	
\end{IEEEproof}

It can be seen that Theorem 1 results into a $4$-variate Fox's H-function. Note that multivariate Fox's H- function is extensively used to analyze wireless systems for complicated fading channels \cite{Bhardwaj2022_multihop, Du2020_RIS_THz_HW_Impaiment}. However, to simplify further, as a specific instance of Theorem 1, we consider the special case where $\eta=1$ and $\kappa=0$, which  leads to the  $\alpha$-$\mu$ model \cite{Papasotiriou2021_scientific_report}. In this context, we provide  statistical results in terms of single-variate Fox's H-function:
\begin{my_corollary} \label{cor:pdf_alpha_mu}
	The PDF and CDF of the THz link with random path loss, antenna misalignment errors, and transceiver hardware impairments with $\alpha$-$\mu$ short-term fading are given by
		\small
\begin{flalign} \label{eq:pdf_hfpl}
	&f_{\gamma}(\gamma) = \frac{\alpha \mu^{\mu} \rho^2 z^k }{ \Omega^{\alpha\mu}\Gamma (\mu)  a_l^{\alpha\mu-z}} \frac{1}{2\big(1-\gamma k_h^2\big)\sqrt{\frac{\bar{\gamma} \gamma}{1-\gamma k_h^2}}} \bigg(\sqrt{\frac{\gamma_h}{\bar{\gamma}}}\bigg)^{\alpha\mu-1} \nonumber \\ & H^{3+k,0}_{2+k,3+k}\Bigg[ \frac{\mu \gamma_h^{\frac{\alpha}{2}}}{\Omega^{\alpha\mu} a_l^\alpha \bar{\gamma}^{\frac{\alpha}{2}}} \Bigg| \begin{matrix}  (1-\alpha\mu+\rho,\alpha)_2, (-\alpha\mu+z+1,\alpha)_k  \\ (0,1), (-\alpha\mu+\rho,\alpha)_2, (-\alpha\mu+z,\alpha)_k	\end{matrix}\Bigg] 
\end{flalign}
\begin{flalign} \label{eq:cdf_hfpl}
	&F_{\gamma}(\gamma) = \frac{\alpha \mu^{\mu} \rho^2 z^k }{ \Omega^{\alpha\mu}\Gamma (\mu)  a_l^{\alpha\mu-z}} \bigg(\sqrt{\frac{\gamma_h}{\bar{\gamma}}}\bigg)^{\alpha\mu}  H^{3+k,1}_{3+k,4+k} \nonumber \\ & \Bigg[ \frac{\mu \gamma_h^{\frac{\alpha}{2}}}{\Omega^{\alpha\mu} a_l^\alpha \bar{\gamma}^{\frac{\alpha}{2}}} \Bigg| \begin{matrix} (1-\alpha\mu,\alpha),  (1-\alpha\mu+\rho,\alpha)_2, (-\alpha\mu+z+1,\alpha)_k  \\ (0,1), (-\alpha\mu+\rho,\alpha)_2,  (-\alpha\mu+z,\alpha)_k, (-\alpha\mu,\alpha)	\end{matrix}\Bigg] 
\end{flalign}
\end{my_corollary}
\normalsize
\begin{IEEEproof}
	The proof follows similar steps as that of Theorem 1.
\end{IEEEproof}

In the following corollary, we  simplify the statistical outcomes by assuming that short-term fading can be disregarded. This scenario might arise in specific situations, such as when dealing with a shorter link. This assumption   allows the representation  of PDF and CDF  in terms of  incomplete Gamma functions.  
\begin{my_corollary} \label{cor:pdf_single}
The PDF and CDF of SNR with the effect of random path loss and antenna misalignment errors with transceiver hardware impairments for the THz link are given by
\begin{flalign} \label{eq:pdf_hl_hp}
	&f_{\gamma}(\gamma) =  -\frac{z^k \rho^2 a_l^{z-\rho} \gamma^{\frac{\rho-1}{2}} (z-\rho)^{-k}}{\Gamma(k) \bar{\gamma_h}^{\frac{\rho-1}{2}}} \frac{1}{2\big(1-\gamma k_h^2\big)\sqrt{\frac{\bar{\gamma} \gamma}{1-\gamma k_h^2}}} \nonumber \\ & \Bigg[ \frac{1}{z-\rho} \Big[\Gamma(k+1) - \Gamma\Big(k+1, (z-\rho) \ln \Big(\frac{a_l \bar{\gamma}}{\gamma_h}\Big)\Big)\Big]  \nonumber \\ &  - \Big[\Gamma(k) - \Gamma\Big(k,(z-\rho) \ln \Big(\frac{a_l \bar{\gamma}}{\gamma_h}\Big)\Big)\Big] \Bigg]
\end{flalign}	

\begin{flalign} \label{eq:cdf_hl_hp}
	& F_{\gamma}(\gamma) = - {z^k \rho^2 a_l^{z} (z-\rho)^{-k}} \Bigg[ \frac{k}{z-\rho} \bigg[\frac{1}{\rho} \Big(\ln\Big(\frac{a_l \bar{\gamma}}{\gamma_h}\Big)\Big)^{-\rho} \nonumber \\ & + \sum_{j=0}^{k} \frac{\Big((z-\rho)\Big)^j}{j!} (z+2\rho)^{-j-1} \Gamma\Big(j+1,(z+2\rho) \ln\Big(\frac{a_l \bar{\gamma}}{\gamma_h}\Big)\Big) \bigg] \nonumber \\ & + \bigg[\frac{1}{\rho} \Big(\ln\Big(\frac{a_l \bar{\gamma}}{\gamma_h}\Big)\Big)^{-\rho} + \sum_{j=0}^{k-1} \frac{\Big((z-\rho)\Big)^j}{j!}(z+2\rho)^{-j-1} \nonumber \\ & \times \Gamma\Big(j+1,(z+2\rho) \ln\Big(\frac{a_l \bar{\gamma}}{\gamma_h}\Big)\Big)\bigg] \Bigg]
\end{flalign}
	
\end{my_corollary}
\begin{IEEEproof}
	The proof is presented in Appendix B.
\end{IEEEproof}
By utilizing the statistical findings presented in Theorem 1, Corollary 1, and Corollary 2, we can derive analytical expressions for various performance metrics in THz wireless systems. In the following subsection, we focus on analyzing the outage probability to showcase the practicality of the proposed generalized model in terms of analytical tractability.
\subsection {Outage Performance Analysis}
The outage probability of a THz system is the probability that the wireless link quality falls below a specified threshold of SNR $\gamma_{\rm th}$. The outage probability is an important performance metric, which can be obtained using the CDF function $F_\gamma(\gamma)=P(\gamma < \gamma_{\rm th}) $. We can substitute $\gamma = \gamma_{\rm th}$ in  \eqref{eq:cdf_aekm_rpl}, \eqref{eq:cdf_hfpl}, and \eqref{eq:cdf_hl_hp} to get the outage probability for the THz transmission link under different propagation scenarios.

An asymptotic expression for the outage probability can be obtained by invoking $\bar{\gamma}\to \infty$  \eqref{eq:cdf_aekm_rpl}, \eqref{eq:cdf_hfpl}, and \eqref{eq:cdf_hl_hp}. As an illustration, the asymptotic outage probability in high SNR region for \eqref{eq:cdf_aekm_rpl} can be derived using the method of residues as described in \cite{AboRahama_2018}, while the asymptotic outage probability for \eqref{eq:cdf_hfpl} can be obtained using the approach described in \cite[Th. 1.11]{Kilbas_2004}. A general expression for the asymptotic outage probability for both the cases is given by
\begin{flalign} \label{eq:outage_fpl_asymp}
	&P_{\rm out}^\infty(\gamma_{\rm th}) = C_1 \bigg(\frac{\gamma_{h_{\rm th}}}{\bar{\gamma}}\bigg)^{\frac{\alpha\mu}{2}} + C_2 \bigg(\frac{\gamma_{h_{\rm th}}}{\bar{\gamma}}\bigg)^{\frac{\rho}{2}} + C_3 \bigg(\frac{\gamma_{h_{\rm th}}}{\bar{\gamma}}\bigg)^{\frac{z}{2}}
\end{flalign}
where $ C_1 $, $ C_2 $, and $ C_3 $ are constants. Analyzing the exponents of the average SNR in \eqref{eq:outage_fpl_asymp}, the diversity order of the considered system can be obtained as
\begin{flalign}
	&DO = \biggl\{\frac{\alpha\mu}{2}, \frac{\rho}{2}, \frac{z}{2}\biggr\}.
\end{flalign}
The diversity order offers multiple options to mitigate the effects of antenna misalignment errors and atmospheric absorption. By understanding the diversity order, we can establish guidelines for effectively utilizing the beam width and link distance to counteract the impact of antenna misalignment errors and random atmospheric absorption. Consequently, the appropriate selection of beam width (to address pointing errors) and link distance (to deal with atmospheric absorption) can help overcome the signal fading.
\begin{figure}[tp]	
	\centering
	\includegraphics[scale=0.26]{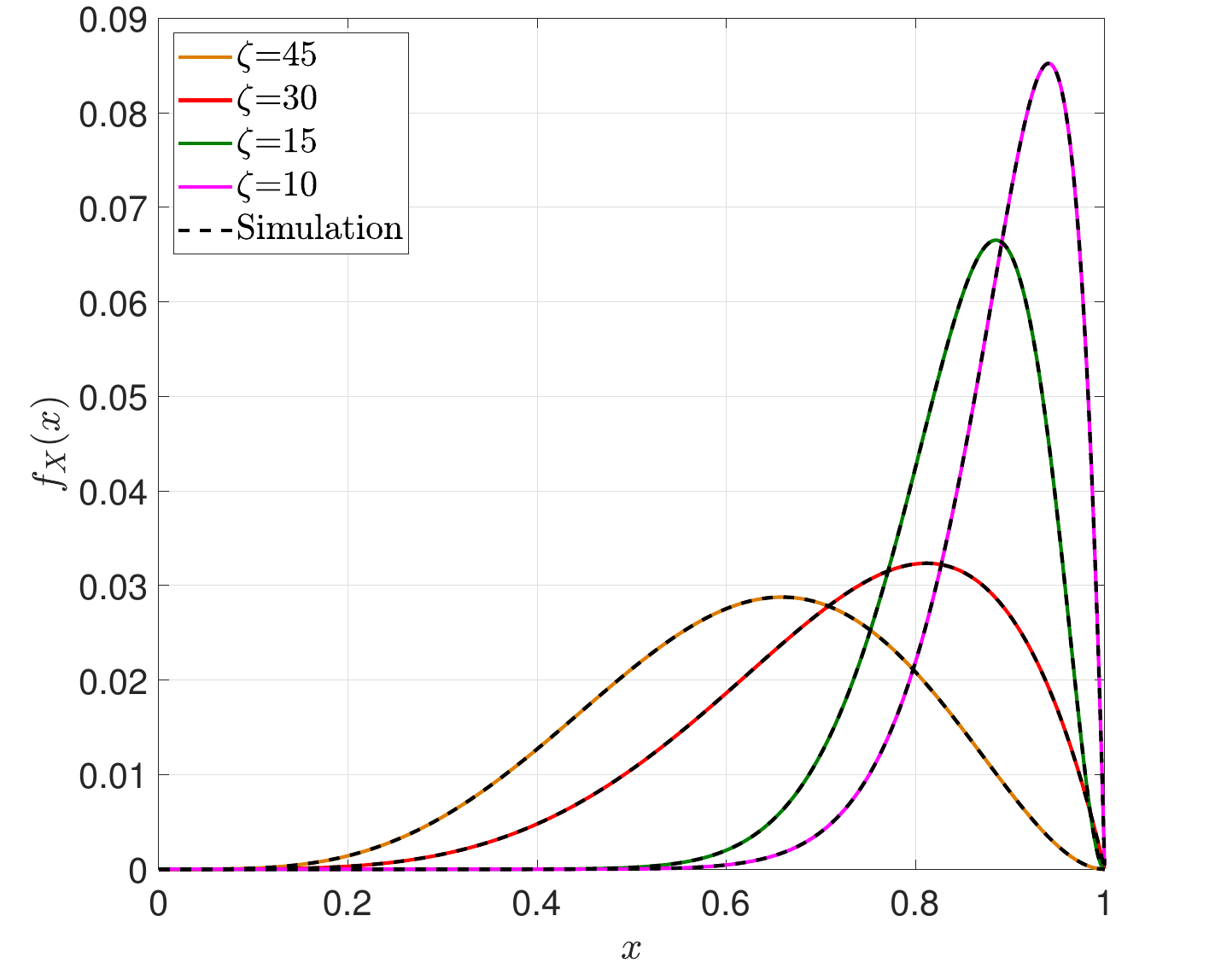}
	\caption{Combined PDF of random path-loss with antenna misalignment errors.}
	\label{fig:pdf}	
\end{figure}
\begin{figure*}[tp]
	\centering
		\subfigure[For different values of $\zeta$ ]{\includegraphics[scale=0.24]{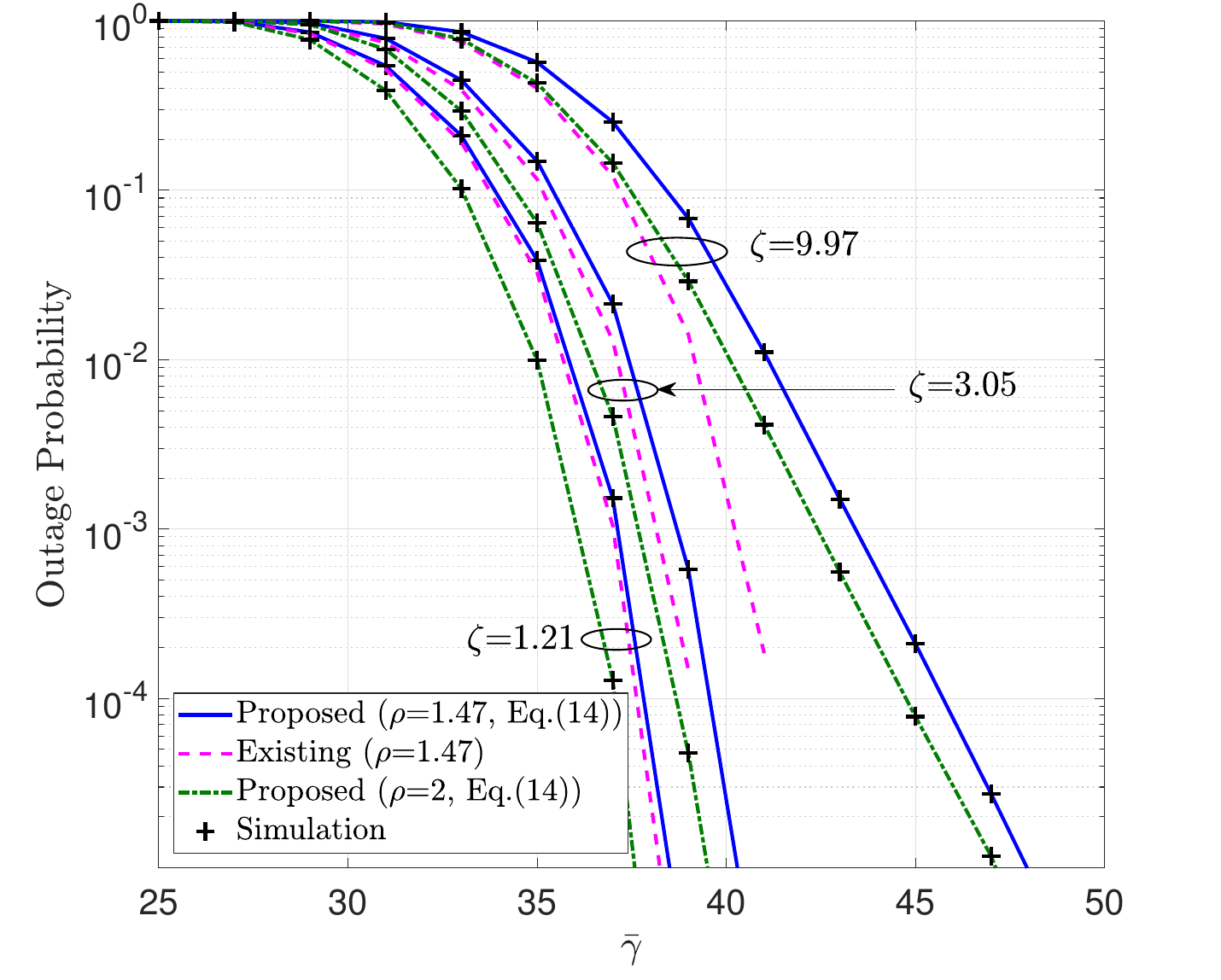}}
		\subfigure[For different values of $\rho$ and $\mu$, $\alpha=1$, $k=1$ ]{\includegraphics[scale=0.24]{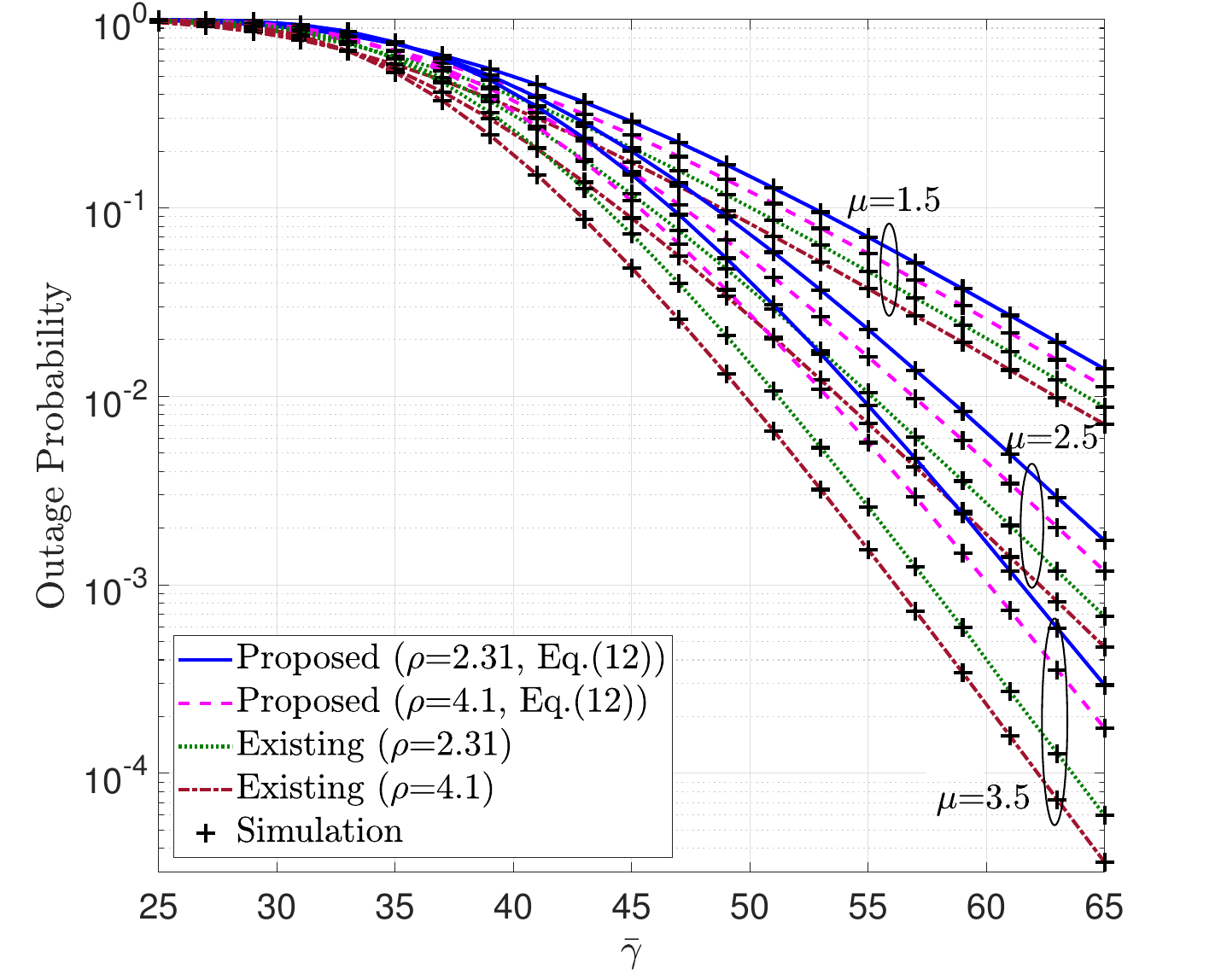}}
			\subfigure[For different values of $\alpha$ and $\kappa$, $\mu=2$, $\eta=1$, $\rho=4.1$, $k=3$ .]{\includegraphics[scale=0.24]{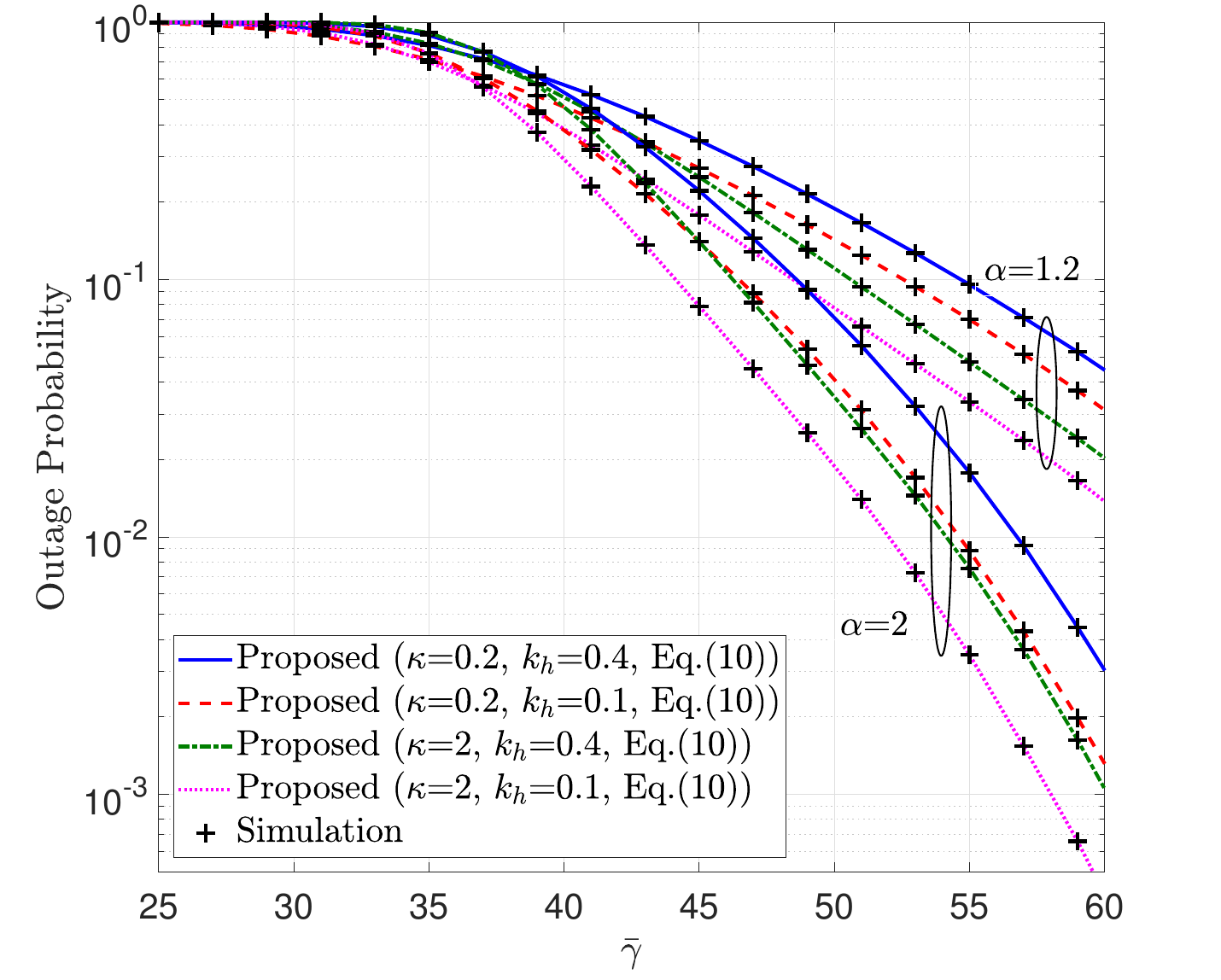}}
		\caption{Outage probability performance.}
		\label{fig:outage}
\end{figure*}

\section{Simulation and Numerical Analysis}
In this section, we assess the performance of the proposed scheme and evaluate its accuracy through simulations and numerical analysis conducted using MATLAB. We utilize Monte Carlo simulations to validate our statistical analysis in the presence of the random path-loss. Our study investigates the THz link performance for numerous channel conditions by adopting different values of  absorption coefficient $\zeta$. The product of shape and scaling parameters for the random path-loss ($k  \beta$) in \eqref{eq:pdf_path_loss} provides the average value of the absorption coefficient $\zeta$ \cite{fog}.

In Fig. \ref{fig:pdf}, we illustrates the PDF of \eqref{eq:pdf_hl_hp} across a wide range of $\zeta$ values, representing different propagation scenarios by only considering the impact of random path-loss and antenna misalignment errors. Observing the plots, it is evident that for smaller absorption coefficient values ($\zeta$), the random variable tends to cluster around the value of $1$. Conversely, with higher $\zeta$ values, the variable attenuates and shows a greater tendency to be situated in regions with lower amplitudes.

Fig. \ref{fig:outage} presents the outage probability performance of the THz link under various channel scenarios. In Fig. \ref{fig:outage}(a), we analyze the outage probability by solely considering random path-loss and antenna misalignment errors, as outlined in {Corollary \ref{cor:pdf_single}}. For this analysis, we exclude the effect of short-term fading and transceiver hardware impairments on the THz link's performance. The plots clearly demonstrate that with increases in $\zeta$ value, the outage probability increases. Additionally, we demonstrate the impact of antenna misalignment errors on the outage probability performance. The system's outage performance improves as the antenna misalignment error parameter $\rho$ increases. We compare our proposed results with existing ones employing a deterministic $\zeta$. Figure \ref{fig:outage}(a) clearly shows that using the deterministic model of $\zeta$ sets an upper bound on the performance of the THz link. At an SNR of $ 45 $ \mbox{dB}, the outage probability is nearly four times lower when employing the random path-loss model compared to the deterministic path-loss model used in previous works. This highlights the importance of considering the random path-loss model to accurately capture the behavior of the actual system. 

Fig. \ref{fig:outage}(b) illustrates the outage performance of the THz link combining the effect of $\alpha$-$\mu$ random path-loss, antenna misalignment error, and short-term fading with antenna misalignment error parameter $\rho$ and channel fading parameter $\mu$ (Corollary \ref{cor:pdf_alpha_mu}). The outage probability decreases as the number of multi-path clusters increases. Additionally, the outage performance improves with an increase in $\rho$, indicating a decrease in antenna misalignment error. It can be observed from the plots that the outage performance improves by approximately $ 30 $ times when $\mu$ is increased from $ 1.5 $ to $ 2.5 $ at an SNR of $ 60 $ \mbox{dB} for $\rho = 4.1$. The figure demonstrates that the slope of the plots remains constant when $\rho$ and $ k $ ($z = 8.686/(\beta d)$) values change but changes only with $\alpha$ and $\mu$, thereby confirming our diversity order analysis.

Fig. \ref{fig:outage}(c) demonstrates the outage performance of the generalized channel model by considering random path-loss, antenna misalignment error, $\alpha$-$\eta$-$\kappa$-$\mu$ short-term fading, and transceiver hardware impairments with parameters $\mu=2$, $\eta=1$, $\rho=4.1$, and $k=3$. The figure reveals that the link's outage performance improves with higher values of the fading parameters $\alpha$ and $\kappa$. Additionally, the impact of transceiver hardware impairment is evident in the outage probability. As the hardware impairment coefficient $k_h$ increases, the outage probability also increases. Notably, when $\alpha$ increases from $1.2$ to $2$ for $\kappa=0.2$ and $k_h=0.4$ at an SNR of $55$ dB, the outage probability is reduced by nearly $9$ times. Due to space constraints, we have provided only a limited set of simulation results. A more comprehensive simulation results will be presented in the transaction version of this paper.

\section{Conclusion}\label{section_conclusion}
In this paper, we developed a generalized channel model for THz wireless system applicable in diverse propagation scenarios at a THz  frequency. We derived the PDF and CDF for the random path-loss by employing the Gamma distribution to model the atmospheric absorption coefficient  We conducted a comprehensive statistical analysis of the THz link, taking into account the combined impact of random path-loss, THz antenna misalignment errors, versatile $\alpha$-$\eta$-$\kappa$-$\mu$ channel fading, and transceiver hardware impairments. By deriving the outage probability and determining the diversity order, we provide more insights into the system's performance. Moreover, the performance analysis of the THz link underscores the importance of considering the random path-loss model for achieving an accurate and realistic representation of THz wireless communication networks. By accounting for the random path-loss, generalized short-term fading and transceiver hardware impairment, the proposed channel model can better capture the actual behavior and performance of the system at true THz frequencies. Validating the proposed model with experimental data would be a possible avenue for future research.

\section*{Appendix A}
To derive the combined PDF of random path-loss, antenna misalignment error, and short-term fading, we first derive the joint PDF of short-term fading and antenna misalignment error $f_{h_{\rm fp}}(x)$, which is given by \cite{papoulis_2002}
\begin{flalign} \label{eq:combined_hfp_eqn}
	f_{h_{fp}}(z) = \int_{z}^{\infty} \frac {1}{x} f_{h_{f}}(x)  f_{h_{p}}\left ({\frac {z}{x}}\right)  \mathrm {d}x.
\end{flalign}

Substituting the respective PDFs of short-term fading and antenna misalignment error from \eqref{eq:pdf_aekm} and \eqref{eq:pdf_pointing_thz} in \eqref{eq:combined_hfp_eqn}, utilizing the Mellin Barnes type integral form of the exponential function and substituting $ \ln\big(\frac{z}{x}\big)=t $, we rewrite the integration and get the inner integral $ \int_{0}^{\infty} {(e^{-t})^{\alpha(1+A_1+A_2+s_1+s_2+s_3+s_4)-\rho+1}}  t \mathrm {d}t $. We solve the inner integral by applying the identity \cite[(3.381/4)]{Gradshteyn} and after some mathematical manipulation and utilizing the definition of multivariate Fox's H-function, we get the joint PDF of short-term fading and antenna misalignment error as
\begin{flalign} \label{eq:pdf_hfp_aekm}
	&f_{h_{fp}}(z) = \frac{ \psi_1 \rho^2 (z)^{\alpha(1+A_1+A_2)}}{(\hat{r}^\alpha)^{1+\frac{\mu}{2}}} \nonumber \\ &\times H^{0,2;1,0;1,1;1,0;1,0}_{2,2;0,1;2,3;1,3;0,1} \Bigg[ \begin{matrix}~ V_3~ \\ ~V_4~ \end{matrix} \Bigg| A_{3}z^{\alpha}, \frac{A_4^2}{4}z^{\alpha},\frac{A_5^2}{4}z^{\alpha},\psi_3z^{\alpha} \Bigg],
\end{flalign}
where $V_3 = \big\{(-A_2;1,0,1,0)\big\}, \big\{\rho-\alpha-\alpha A_1 -\alpha A_2; \alpha, \alpha, \alpha, \alpha\big\}: \big\{(-,-)\big\} ; \big\{(-A_1,1)(\frac{1}{2},1)\big\} ;\big\{(\frac{1}{2},1)\big\}; \big\{-, -\big\} $ and $V_4 = \big\{(-1-A_1 -A_2;1,1,1,0) \big\}, \big\{\rho-1-\alpha-\alpha A_1 -\alpha A_2; \alpha, \alpha, \alpha, \alpha\big\} : \big\{(0,1) \big\} ; \big\{(0,1),(-A_1,1),(\frac{1}{2},1) \big\} ; \big\{(0,1),(-A_2,1),(\frac{1}{2},1) \big\} ; \\ \big\{ (0,1)\big\}$.

Now, we need to multiply the joint PDF derived in \eqref{eq:pdf_hfp_aekm}, with the PDF of random path-loss in \eqref{eq:pdf_path_loss} to derive the combined PDF of short-term fading, antenna misalignment error, and random path-loss. The combined PDF equation is given by \cite{papoulis_2002}
\begin{flalign} \label{eq:combined_hfpl_eqn_aekm}
	f_{h_{fpl}}(y) = \int_{\frac{y}{a}}^{\infty} \frac {1}{x} f_{h_{fp}}(x) f_{h_{l}}\left ({\frac {y}{x}}\right) \mathrm {d}x.
\end{flalign}

Similarly, plugging the PDFs of \eqref{eq:pdf_path_loss} and \eqref{eq:pdf_hfp_aekm} in \eqref{eq:combined_hfpl_eqn_aekm} and substituting $ \ln\big(\frac{ax}{y}\big) = t $, and following the similar procedure, we get the combined PDF of fading, antenna misalignment error and random path-loss in \eqref{eq:pdf_aekm_rpl}. The PDF of SNR can be derived by simple transformation of random variable \cite{papoulis_2002} as $f_{\gamma}(\gamma) = \frac{1}{2\sqrt{\gamma \gamma_0}}f_{h_{fpl}}\Big(\sqrt{\frac{\gamma}{\gamma_0}}\Big)$. To derive the CDF, we integrate the PDF $ F_{h_{fpl}}(y) = \int_{0}^{y} f_{h_{fpl}}(y) dy $, to get the inner integral as $ \int_{0}^{y} y^{\alpha(1+A_1+A_2+s_1+s_2+s_3+s_4)} dy $. Solving the inner integral and applying the definition of Fox's H-function, the CDF is given in \eqref{eq:cdf_aekm_rpl}. Similar to the PDF, the CDF of the SNR can be derived by simple transformation of a random variable as $F_{\gamma}(\gamma) = \Big(\sqrt{\frac{\gamma}{\gamma_0}}\Big)$ to conclude the proof.

\section*{Appendix B}
The combined PDF of random path loss and antenna misalignment error $f_{h_{\rm lp}}(x)$ is given by \cite{papoulis_2002}
\begin{flalign} \label{eq:combined_hl_hp_eqn}
	f_{h_{\rm lp}}(x) = \int _{x}^{a} \frac {1}{h_l} f_{h_{l}}(h_l)  f_{h_{p}}\left ({\frac {x}{h_l}}\right)  \mathrm {d}h_l.
\end{flalign}

Substituting the PDFs of the random path-loss and antenna misalignment error from  \eqref{eq:pdf_path_loss} and \eqref{eq:pdf_pointing_thz}, respectively, in \eqref{eq:combined_hl_hp_eqn}, to get the PDF as

\begin{flalign} \label{eq:combined_hl_hp_pdf_der}
	f_{h_{lp}}(x) = - \frac{z^k \rho^2 x^{\rho-1}}{\Gamma(k)} \int _{x}^{a_l} h_{l}^{z-\rho -1} \bigg[ln\bigg(\frac{a_l}{h_{l}}\bigg)\bigg]^{k-1}    \ln\bigg(\frac{x}{h_l}\bigg)   \mathrm {d}h_l
\end{flalign}
substituting $\ln(\frac{a_l}{h_l}) = t$ in \eqref{eq:combined_hl_hp_pdf_der}, and applying the identity $\int_{0}^{u} x^{\nu-1} e^{-\mu x} dx = \mu^{-\nu}\big[\Gamma(\nu)-\Gamma(\nu,\mu u)\big]$ in \eqref{eq:combined_hl_hp_pdf_der}, we get the combined PDF of random path-loss and antenna misalignment error in \eqref{eq:pdf_hl_hp}. The PDF of SNR can be derived by simple transformation of a random variable \cite{papoulis_2002} as $f_{\gamma}(\gamma) = \frac{1}{2\sqrt{\gamma \gamma_0}}f_{h_{fpl}}\Big(\sqrt{\frac{\gamma}{\gamma_0}}\Big)$. To derive the CDF, we will integrate the PDF $F_{h_{lpl}}(y) = \int_{0}^{y} f_{h_{lpl}}(x) dx $ and use the series expansion of the upper incomplete Gamma function $\Gamma(a,z) = (a-1)! e^{-z}\sum_{j=0}^{a-1}\frac{z^j}{j!}$, to get

\small
\begin{flalign} %\label{eq:combined_hl_hf_eqn} 
	& F_{h_{lp}}(y) = - \frac{z^k \rho^2 a_l^{z-\rho}   (z-\rho)^{-k}}{\Gamma(k)} \nonumber \\ & \Bigg[ \frac{1}{z-\rho} \Big[ \int_{0}^{y}  x^{\rho-1} \Gamma(k+1)  dx \nonumber \\ &  -  k! \sum_{j=0}^{k} \frac{\Big((z-\rho)\Big)^j}{j!}  \int_{0}^{y}  x^{\rho-1}  \Big(\frac{a_l}{x}\Big)^{-(z-\rho)} \Big(\ln\big(\frac{a_l}{x}\big)\Big)^j   dx  \Big]  \nonumber \\ &  - \Big[ \int_{0}^{y}  x^{\rho-1} \Gamma(k) dx  - (k-1)! \nonumber \\ &  \sum_{j=0}^{k-1} \frac{\Big((z-\rho)\Big)^j}{j!} \int_{0}^{y}  x^{\rho-1}   \Big(\frac{a_l}{x}\Big)^{-(z-\rho)} \Big(\ln\big(\frac{a_l}{x}\big)\Big)^j  dx  \Big] \Bigg]
\end{flalign} \normalsize
substituting $\ln(\frac{a_l}{x}) = t$ and applying the identity \cite[3.351,2]{Gradshteyn} $\int_{u}^{\infty} x^n e^{-\mu x} dx = \mu^{-n-1} \Gamma(n+1, \mu u)$, we get the combined CDF of random path-loss and antenna misalignment error in \eqref{eq:cdf_hl_hp}. The CDF of the SNR can be derived by transforming the random variable as $F_{\gamma}(\gamma) = \Big(\sqrt{\frac{\gamma}{\gamma_0}}\Big)$ to finish the proof.
	
\bibliographystyle{IEEEtran}
\bibliography{jsac_bib_file}
\end{document}